\documentclass[english,aps,preprint]{revtex4-1}
\usepackage[T1]{fontenc}
\usepackage[latin9]{inputenc}
\usepackage{amsmath}
\usepackage{amssymb}

\makeatletter

\providecommand{\LyX}{L\kern-.1667em\lower.25em\hbox{Y}\kern-.125emX\@}

\@ifundefined{textcolor}{}
{%
 \definecolor{BLACK}{gray}{0}
 \definecolor{WHITE}{gray}{1}
 \definecolor{RED}{rgb}{1,0,0}
 \definecolor{GREEN}{rgb}{0,1,0}
 \definecolor{BLUE}{rgb}{0,0,1}
 \definecolor{CYAN}{cmyk}{1,0,0,0}
 \definecolor{MAGENTA}{cmyk}{0,1,0,0}
 \definecolor{YELLOW}{cmyk}{0,0,1,0}
 }

\makeatother

\usepackage{babel}

\begin{document}

\title{Scaling dimensions in hidden Kerr/CFT}

\author{David A. Lowe, Ilies Messamah and Antun Skanata}

\email{lowe, ilies_messamah, antun_skanata@brown.edu}

\affiliation{Department of Physics, Box 1843, Brown University, Providence, RI
02912, USA}
\begin{abstract}
It has been proposed that a hidden conformal field theory (CFT) governs
the dynamics of low frequency scattering in a general Kerr black hole
background. We further investigate this correspondence by mapping
higher order corrections to the massless wave equations in a Kerr
background to an expansion within the CFT in terms of higher dimension
operators. This implies the presence of infinite towers of CFT primary
operators with positive conformal dimensions compatible with unitarity.
The exact Kerr background softly breaks the conformal symmetry and
the scaling dimensions of these operators run with frequency. The
scale-invariant fixed point is dual to a degenerate case of flat spacetime.
\end{abstract}
\maketitle

\section{introduction}

Several years ago an intriguing conjecture was made that quantum gravity
around a general Kerr black hole background is dual to a conformal
field theory \cite{Guica:2008mu}. Most of the evidence for this conjecture
has been established for the case of extremal Kerr. For example, the
properties of near-super radiant modes in extremal Kerr could be explained
via a dual two-dimensional conformal field theory \cite{Bredberg:2009pv}.

The conjecture is surprising because the general Kerr black hole does
not have any obvious geometric symmetry near its horizon that might
explain the conformal structure. More recently, it was observed that
the limit of low-frequency scattering in the near-region of a black
hole does possess such a hidden conformal structure \cite{Castro:2010fd}.
This was observed by studying the massless scalar wave equation in
the general Kerr background.

If this hidden CFT viewpoint can be put on a sound footing, these
techniques would lead to a radically new way to treat the quantum
physics of the entire class of Kerr black holes, including the Schwarzschild
limit. In addition to accounting for the quantum entropy of the black
hole, it would provide an efficient mechanism for computing of scattering
(at least in the near-region) in a small frequency expansion. Moreover
if the central charge can be computed in a reliable way from the gravity
side, this proposal would yield dramatic new insight into the physics
of the black hole microstates that account for the Bekenstein-Hawking
entropy.

In the present work we study this hidden conformal symmetry in more
detail, and further develop the correspondence between CFT primaries
and bulk fields. We find that the dual CFT must contain infinite towers
of quasi-primary operators with positive conformal weights, compatible
with unitarity. However the full Kerr geometry softly breaks the conformal
symmetry, and induces a nontrivial running of the scaling dimensions
of these operators. The fixed point where the hidden conformal symmetry
becomes exact is flat spacetime. This indicates that if there is an
exact CFT underlying the dynamics of Kerr, there is not a smooth geometrical
limit connecting the low frequency limit of general Kerr, with the
promising studies of dynamics of extremal Kerr.

\section{Review of Hidden Kerr/CFT}

The Kerr metric in Boyer-Lindquist coordinates (with $c=G=1)$ is\begin{eqnarray*}
ds^{2} & = & (1-2Mr/\Sigma)dt^{2}+\left(4Mar\sin^{2}\left(\theta\right)/\Sigma\right)dtd\phi-\left(\Sigma/\Delta\right)dr^{2}-\Sigma d\theta^{2}-\\
 &  & \sin^{2}\left(\theta\right)\left(r^{2}+a^{2}+2Ma^{2}r\sin^{2}\left(\theta\right)/\Sigma\right)d\phi^{2}\end{eqnarray*}
where $M$ is the black hole mass, $J=aM$ is the angular momentum,
$\Sigma=r^{2}+a^{2}\cos^{2}\theta$ and $\Delta=r^{2}-2Mr+a^{2}$.
The outer and inner horizons sit at $r_{\pm}=M\pm\sqrt{M^{2}-a^{2}}$.
It will be convenient later to define left/right temperatures\[
T_{L}=\frac{M^{2}}{2\pi J}\,,\quad T_{R}=\frac{\sqrt{M^{4}-J^{2}}}{2\pi J}\,.\]
Let us consider the wave equations for massless fields in this background.
Teukolsky \cite{Teukolsky:1972my} found that it was possible to separate
the wave equation in this background for general spin massless fields.
Decomposed into spin-weighted spheroidal harmonics, with spin-weight
$s$ (see \cite{Teukolsky:1972my} for details) the solutions can
be written\begin{equation}
\psi_{s}=e^{-i\omega t}e^{im\phi}S(\theta)R(r)\label{eq:sepsol}\end{equation}
The angular equation takes the form\begin{equation}
\frac{1}{\sin\theta}\frac{d}{d\theta}\left(\sin\theta\frac{dS}{d\theta}\right)+\left(a^{2}\omega^{2}\cos^{2}\theta-\frac{m^{2}}{\sin^{2}\theta}-2a\omega s\cos\theta-\frac{2ms\cos\theta}{\sin^{2}\theta}-s^{2}\cot^{2}\theta+E-s^{2}\right)S=0\label{eq:angular}\end{equation}
where $E$ is the separation constant. The eigenvalue $E$ is constrained
by the requirement that $S$ be regular at $\theta=0,\pi$. For the
special case $a\omega=0$ this may be computed exactly $E=\ell(\ell+1)$.
For general $\omega$ this may be computed numerically, or as a series
expansion. The radial equation takes the form\begin{eqnarray}
\Delta^{-s}\frac{d}{dr}\left(\Delta^{s+1}\frac{dR}{dr}\right)+\Biggl(\Bigl[\left(r^{2}+a^{2}\right)^{2}\omega^{2}-4aMr\omega m+a^{2}m^{2}+\nonumber \\
2ia(r-M)ms-2iM(r^{2}-a^{2})\omega s\Bigr]\Delta^{-1}+2ir\omega s-E+s(s+1)-a^{2}\omega^{2}\Biggr)R & = & 0\,.\label{eq:radial}\end{eqnarray}
The relation between solutions $\psi_{s}$ and canonically normalized
massless fields (components of the field strength for spin 1, and
components of the Weyl tensor for spin 2) are given by (see \cite{Teukolsky:1972my}
for further details)\begin{eqnarray}
\mathrm{scalar}\,\Phi & = & \psi_{0}\nonumber \\
\mathrm{vector}\,\varphi_{0} & = & \psi_{1}\qquad\varphi_{2}=\chi^{2}\psi_{-1}\nonumber \\
\mathrm{tensor}\,\Psi_{0} & = & \psi_{2}\qquad\Psi_{4}=\chi^{4}\psi_{-2}\label{eq:teukmodes}\end{eqnarray}
where $\chi=-1/\left(r-ia\cos\theta\right)$. 

Castro, Maloney and Strominger \cite{Castro:2010fd} considered the
scalar case of the wave equation ($s=0)$ and noticed that in a low
frequency expansion $\omega M\ll1$, that the leading order term in
the radial equation in the near-region, where $\omega r\ll$1, reduces
to a hypergeometric equation. They then showed that the full solution
in this limit transformed as a representation of $SL(2,\mathbb{R})\times SL(2,\mathbb{R})$
(broken to $U(1)\times U(1)$ when the periodic identification of
$\phi\sim\phi+2\pi$ is taken into account). This led them to propose
a hidden Kerr/CFT duality, with a scalar mode with angular momentum
$\ell$ being identified with a CFT operator of conformal weight $(h_{L},h_{R})=(\ell,\ell)$.
If one further speculates that the hidden $SL(2,\mathbb{R})\times SL(2,\mathbb{R})$
extends to a left-right Virasoro algebra with central charges $(c_{L},c_{R})=12J$
then the Cardy formula for the CFT entropy agrees exactly with the
Kerr horizon entropy $S=\mathrm{Area}/4$.

In the following we will generalize CMS \cite{Castro:2010fd} by showing
that the entire set of higher order frequency corrections can be organized
into a CFT-like expansion. The precise statement is that the scaling
dimensions run with frequency, which implies the CFT is deformed away
from its exact conformal fixed point. Unfortunately we will see that
the exact fixed point is dual to the $M=0$ solution (i.e. flat spacetime)
which seems to be a degenerate limit at odds with the $(c_{L},c_{R})=12J$
proposal of \cite{Castro:2010fd}.

\section{Series Solutions to the Teukolsky equation}

The strategy for finding exact solutions to the equations \eqref{eq:angular}
and \eqref{eq:radial} will be to perform a small frequency ($\omega)$
expansion. Let us begin by considering the angular equation \eqref{eq:angular}.
As shown in \cite{Fackerell1977}, the solution in a small $\omega$
expansion is written in terms of an infinite series of Jacobi polynomials
$P_{j}^{(\alpha,\beta)}(x)$\[
S=e^{a\omega x}\left(\frac{1-x}{2}\right)^{|m+s|/2}\left(\frac{1+x}{2}\right)^{|m-s|/2}\,_{S}U_{lm}(x)\]
where $x=\cos\theta$ and $U$ is \begin{equation}
U=\sum_{j=0}^{\infty}c_{j}P_{j}^{(|m+s|,|m-s|)}(x)\,.\label{eq:jacobiexpan}\end{equation}
Inserting \eqref{eq:jacobiexpan} into \eqref{eq:angular} leads to
a 3-term recurrence relation for the $c_{j}$. The expansion for $U$
is well-defined (in the sense that each $c_{j}$ is determined and
finite) and convergent, provided the separation constant $E$ satisfies
an equation, that may be expressed as a continued fraction using the
recurrence relations. This transcendental equation may then be readily
solved as a power series expansion in $a\omega$, with the result\begin{equation}
E=\ell(\ell+1)-\frac{2s^{2}ma\omega}{\ell(\ell+1)}+\mathcal{O}((a\omega)^{2})\,.\label{eq:eexact}\end{equation}

The radial equation \eqref{eq:radial} may be tackled in a similar
way as studied in \cite{Mano:1996gn,Mano:1996vt}. This time, $R(r)$
is expressed as a series of hypergeometric functions. Defining a rescaled
radial coordinate $\rho=\omega(r_{+}-r)/\epsilon\kappa$, and the
constants $\epsilon=2M\omega$, $\kappa=\sqrt{1-(a/M)^{2}}$ and $\tau=(\epsilon-ma/M)/\kappa$,
the radial function is factored as \begin{equation}
R_{s}(\rho)=e^{i\epsilon\kappa\rho}(-\rho)^{-s-\frac{i}{2}(\epsilon+\tau)}(1-\rho)^{\frac{i}{2}(\epsilon-\tau)}P(\rho)\,.\label{eq:radialfun}\end{equation}
The function $P(\rho)$ then admits the series expansion\begin{equation}
P(\rho)=\sum_{n=-\infty}^{\infty}a_{n}\,_{2}F_{1}(n+\nu+1-i\tau,-n-\nu-i\tau;1-s-i\epsilon-i\tau;\rho)\label{eq:expan}\end{equation}
where the coefficients $a_{n}$ satisfy a three-term linear recursion
relation$ $\begin{equation}
\alpha_{n}a_{n+1}+\beta_{n}a_{n}+\gamma_{n}a_{n-1}=0\label{eq:recurr}\end{equation}
with\begin{eqnarray*}
\alpha_{n} & = & \frac{i\epsilon\kappa(n+\nu+1+s+i\epsilon)(n+\nu+1+s-i\epsilon)(n+\nu+1+i\tau)}{(n+\nu+1)(2n+2\nu+3)}\\
\beta_{n} & = & -E+2\epsilon^{2}-q^{2}\epsilon^{2}/4+(n+\nu)(n+\nu+1)+\frac{\epsilon(\epsilon-mq)(s^{2}+\epsilon^{2})}{(n+\nu)(n+\nu+1)}\\
\gamma_{n} & = & -\frac{i\epsilon\kappa(n+\nu-s+i\epsilon)(n+\nu-s-i\epsilon)(n+\nu-i\tau)}{(n+\nu)(2n+2\nu-1)}\,.\end{eqnarray*}

There exist standard methods for solving such recursion relations,
as discussed in \cite{Gautschi1967}. The general solution can be
expressed as a linear combination of two independent solutions (since,
for example, one can choose arbitrary initial values for $a_{0}$
and $a_{1}$. One solution has coefficients that diverge as $|n|\to\infty$
and is called the dominant solution. The other solution, of most interest
for the present work, is the minimal solution, where the $a_{n}$
converge (or at least diverge less rapidly) at large $|n|$. This
solution must be obtained by tuning $a_{1}$ with respect to $a_{0}$. 

Furthermore, if the $a_{n}$ converge, a continued fraction equation
may be set up to determine the value of the eigenvalue $\nu$. This
may be arranged by solving for $\nu$ in two different ways: by setting
$a_{0}=1$ and evolving the minimal solution to $n=\infty$ or by
evolving the minimal solution to $n=-\infty$. To see this we define
the ratios \[
R_{n}=\frac{a_{n}}{a_{n-1}}\,,\qquad L_{n}=\frac{a_{n}}{a_{n+1}}\]
so that $R_{n}$ converges as $n\to\infty$ and $L_{n}$ converges
as $n\to-\infty$. Then, the three-term recurrence relation \eqref{eq:recurr}
may be rewritten \[
R_{n}=-\frac{\gamma_{n}}{\beta_{n}+\alpha_{n}R_{n+1}}\,,\qquad L_{n}=-\frac{\alpha_{n}}{\beta_{n}+\gamma_{n}L_{n-1}}\]
which may then be developed as convergent continued fractions. These
continued fractions yield the equation\[
R_{1}L_{0}=1\]
which generates a transcendental equation for $\nu$. This may be
solved as a low frequency expansion in $\epsilon$ together with the
coefficients $a_{n}$ as shown in \cite{Mano:1996gn}. This yields
the solution to the Teukolsky equation infalling on the future outer
horizon. The leading terms in the expansion for $\nu$ are \begin{equation}
\nu=\ell-\frac{\epsilon^{2}}{2\ell+1}\left(2+\frac{s^{2}}{\ell(\ell+1)}+\frac{\left(\ell^{2}-s^{2}\right)^{2}}{(2\ell-1)2\ell(2\ell+1)}-\frac{\left(\left(\ell+1\right)^{2}-s^{2}\right)^{2}}{(2\ell+1)(2\ell+2)(2\ell+3)}\right)+\mathcal{O}(\epsilon^{3})\,.\label{eq:nueq}\end{equation}
Interestingly, the expansion \eqref{eq:expan} converges for all finite
$r$ \cite{Mano:1996gn}, so the near-region condition of CMS $\omega r\ll1$
turns out not to be needed. 

The solution to Teukolsky equation outgoing on the future outer horizon
is obtained from the solution to \eqref{eq:radialfun} as\[
R_{out}(\rho)=\Delta(\rho)^{-s}(R_{-s}(\rho))^{*}\,.\]

\section{Bulk Field/ CFT Operator Map}

In a low frequency expansion, the exact solution \eqref{eq:radialfun}
may be expanded as a regular series in $\epsilon$. The leading term
is\begin{equation}
R_{s}^{0}(\rho)=e^{i\epsilon\kappa\rho}(-\rho)^{-s-\frac{i}{2}(\epsilon+\tau)}(1-\rho)^{\frac{i}{2}(\epsilon-\tau)}\,_{2}F_{1}(\nu+1-i\tau,-\nu-i\tau;1-s-i\epsilon-i\tau;\rho)\,.\label{eq:lowfsol}\end{equation}
We may perform a transformation $\rho\to\frac{\rho}{\rho-1}$ in the
argument of the hypergeometric function to give\[
R_{s}^{0}(\rho)=e^{i\epsilon\kappa\rho}(-\rho)^{-s-\frac{i}{2}(\epsilon+\tau)}(1-\rho)^{\frac{i}{2}(\epsilon+\tau)-\nu-1}\,_{2}F_{1}(\nu+1-i\tau,1-s-i\epsilon+\nu;1-s-i\epsilon-i\tau;\frac{\rho}{\rho-1})\]
which agrees with eqn. (6.1) in \cite{Castro:2010fd} (note a typo
in the argument of the hypergeometric function in the arxiv version
of \cite{Castro:2010fd}, corrected in the published version), upon
replacing $\nu$ with its low frequency limit $\ell$, setting $s=0$,
and dropping the first factor, as appropriate for the near-region. 

The argument of \cite{Castro:2010fd} proceeds by noting that \eqref{eq:lowfsol}
solves the equation\[
\mathcal{H}^{2}\psi_{0}=\bar{\mathcal{H}^{2}}\psi_{0}=\ell(\ell+1)\psi_{0}\]
where $\mathcal{H}^{2}$ and $\bar{\mathcal{H}}^{2}$ are the Casimir
operators of the $SL(2,\mathbb{R})\times SL(2,\mathbb{R})$ algebra
generated by\begin{eqnarray}
H_{1} & = & ie^{-2\pi T_{R}\phi}\left(\Delta^{1/2}\partial_{r}+\frac{1}{2\pi T_{R}}\frac{r-M}{\Delta^{1/2}}\partial_{\phi}+\frac{2T_{L}}{T_{R}}\frac{Mr-a^{2}}{\Delta^{1/2}}\partial_{t}\right)\nonumber \\
H_{0} & = & \frac{i}{2\pi T_{R}}\partial_{\phi}+2iM\frac{T_{L}}{T_{R}}\partial_{t}\nonumber \\
H_{-1} & = & ie^{2\pi T_{R}\phi}\left(-\Delta^{1/2}\partial_{r}+\frac{1}{2\pi T_{R}}\frac{r-M}{\Delta^{1/2}}\partial_{\phi}+\frac{2T_{L}}{T_{R}}\frac{Mr-a^{2}}{\Delta^{1/2}}\partial_{t}\right)\label{eq:hgens}\end{eqnarray}
and\begin{eqnarray}
\bar{H}_{1} & = & ie^{-2\pi T_{L}\phi+\frac{t}{2M}}\left(\Delta^{1/2}\partial_{r}-\frac{a}{\Delta^{1/2}}\partial_{\phi}-2M\frac{r}{\Delta^{1/2}}\partial_{t}\right)\nonumber \\
\bar{H}_{0} & = & -2iM\partial_{t}\nonumber \\
\bar{H}_{-1} & = & ie^{2\pi T_{L}\phi-\frac{t}{2M}}\left(-\Delta^{1/2}\partial_{r}-\frac{a}{\Delta^{1/2}}\partial_{\phi}-2M\frac{r}{\Delta^{1/2}}\partial_{t}\right)\label{eq:hbars}\end{eqnarray}
which obey\begin{equation}
[H_{0},H_{\pm1}]=\mp iH_{\pm1}\,,\qquad[H_{-1},H_{1}]=-2iH_{0}\label{eq:sltwo}\end{equation}
and likewise for the others. The conjecture of \cite{Castro:2010fd}
is that this extends to a full left-right Virasoro algebra, with central
charge $c_{L}=c_{R}=12J=12a/M$, and that the conformal weights of
the field $\psi_{0}$ are $(\ell,\ell)$.

For convenience, we identify $L_{n}=-iH_{n}$ and $\bar{L}_{n}=-i\bar{H}_{n}$
so that the $L_{n}$'s satisfy the standard form of the Witt algebra\[
[L_{n},L_{m}]=(n-m)L_{n+m}\,.\]
We begin by investigating bulk modes that satisfy a lowest weight
condition, which should be dual to primary operators in the CFT. Imposing
the equations $L_{1}\psi(r,t,\phi)=\bar{L}_{1}\psi(r,t,\phi)=0$ yields
the solution\[
\psi(r,t,\phi)\propto\left(rr_{+}-a^{2}\right)^{-iam/r_{+}}e^{im\phi-i\omega t}\]
and the condition \begin{equation}
\omega=am/\left(2Mr_{+}\right)\,.\label{eq:prifreq}\end{equation}
The conformal weights are\[
(h_{L},h_{R})=(\frac{iam}{r_{+}},\frac{iam}{r_{+}})\,.\]
So this will solve the scalar field equation of motion in Kerr if
we further identify the Casimir with $\ell(\ell+1)$\begin{equation}
\left(L_{0}^{2}-\frac{1}{2}\left(L_{1}L_{-1}+L_{-1}L_{1}\right)\right)\psi(r,t,\phi)=\ell(\ell+1)\psi(r,t,\phi)\label{eq:zerothorder}\end{equation}
which implies $h_{L}(h_{L}-1)=\ell(\ell+1)$ and $h_{L}=\ell+1$ for
the positive solution. (Here we disagree with the $(\ell,\ell)$ conformal
weight assignment of \cite{Castro:2010fd}).

This leads us to the unfamiliar situation, where the eigenvalue $m$
must be analytically continued to imaginary values to construct a
mode dual to a primary CFT operator. However this is not entirely
unexpected, since the space of infalling modes is a superset containing
the quasi-normal modes of Kerr, studied, for example in \cite{Leaver:1985ax}.
Likewise, the case of quasi-normal modes of the 3d black hole have
been studied in \cite{Birmingham:2001pj}. These modes have complex
eigenvalues for $\omega$ so it is perhaps not too surprising we also
wind up with complex eigenvalues for $m$ prior to imposing periodicity
of $\phi$. This phenomena is encountered in a similar context in
\cite{Chen:2010ik}.

It should also be noted that the frequency condition for a primary
field \eqref{eq:prifreq} becomes\begin{equation}
\omega=\frac{\ell+1}{2iM}\label{eq:prifreqell}\end{equation}
which again takes us out of the low frequency limit $\omega M\ll1$.
We comment further on this point below.

The inner product of these primaries is rather different from the
usual Klein-Gordon norm in the Kerr background. The inner product
of the CFT must yield conjugation that switches $L_{1}\leftrightarrow L_{-1}$
and leaves $L_{0}$ invariant. This is accomplished by Hermitian conjugation,
followed by $\phi\to-\phi$ and $t\to-t$. This suggests the symmetry
may be interpreted directly as acting in an analytic continuation
of the Kerr geometry where $\phi\to i\phi$ and $t\to it$.

We conclude that the large $r$ falloff $r^{-(h_{L}+h_{R})/2}$ of
a mode allows us to read off the conformal weight of the dual CFT
operator $\Delta=h_{L}+h_{R}$. It is worth mentioning that in the
usual AdS/CFT correspondence, the radial fall-off in Poincare coordinates
is instead of the form $\tilde{r}^{-(h_{L}+h_{R})}$. Thus if one
inferred some effective AdS metric from the Kerr Laplacian in the
near region, the relation between coordinates is of the form $r\sim\tilde{r}^{2}$
at large $r$.

From \eqref{eq:lowfsol} we can generalize the above to the higher
$s$ fields, and include the higher powers of $M\omega$ on the right
hand side of \eqref{eq:zerothorder} using the expansion \eqref{eq:eexact}.
In the near region, the large $r$ falloff of \eqref{eq:lowfsol}
takes the form $r^{-s-\nu-1}$. To extract the behavior of the primary
field we must also take into account the normalization of the component
vectors used to set up the Teukolsky equation, that appear in the
definition of the quantities \eqref{eq:teukmodes} as described in
\cite{Teukolsky:1972my}. Thus we extract the large $r$ behavior
of the vector potential $A_{\mu}$ for spin 1, and the behavior of
the graviton $g_{\mu\nu}$ for spin 2, in asymptotically Minkowski
coordinates. This leads us to identify the conformal weight\[
\Delta=h_{L}+h_{R}=-2s+2\nu+2\,.\]
Here we assumed $s\geq0$. For $s<0$ we must recall the canonically
normalized modes are non-trivially related to the solutions of the
Teukolsky equation by \eqref{eq:teukmodes}. So for $s<0$, the large
$r$ falloff of the canonical modes takes the form $r^{s-\nu-1}$
so in general we obtain \begin{equation}
\Delta=-2|s|+2\nu+2\label{eq:condim}\end{equation}
for the conformal weight of a higher spin mode. Thus, at leading order
in $M\omega$, all the massless bulk fields have $\Delta=2$ for the
lowest nontrivial modes of angular momentum.

There are now a number of puzzles we need to address. The $\bar{L}_{n}$
generators act on functions that may be written using the basis $R(r)e^{im\phi-i\omega t}$
but as we see from the exponential prefactors in \eqref{eq:hbars},
the $\bar{L}_{1}$ and $\bar{L}_{-1}$ generators shift the $m$ and
$\omega$ eigenvalues by imaginary amounts. The shift in $\omega$
means the low frequency approximation leading to \eqref{eq:lowfsol}
can no longer be trusted, \emph{so that the whole $SL(2,\mathbb{R})$
associated with these generators is strongly broken down to the $U(1)$
subgroup generated by $\bar{L}_{0}$}. 

This is not the case for the $SL(2,\mathbb{R})$ generated by the
$L_{n}$ because these only shift $m$ by imaginary amounts, and no
small $m$ approximation was used. However as pointed out in \eqref{eq:prifreqell},
the primary modes with respect to $L_{n}$ do take us out of the low
frequency limit. We may still use the $SL(2,\mathbb{R})\times SL(2,\mathbb{R})$
representations of the leading order wave equation to organize the
expansion of the higher order corrections. A priori we have no reason
to expect convergence when we relate the associated CFT operators
with bulk operators, but nevertheless, the mode function expansion
\eqref{eq:expan} happens to converge for all finite $r$.

Of course, as noted in \cite{Castro:2010fd} $SL(2,\mathbb{R})\times SL(2,\mathbb{R})$
is explicitly broken once $\phi$ is periodically identified, which
projects out all noninteger $m$ modes. So in this way, we see how
this can be a low frequency symmetry of the Kerr modes prior to periodic
identification, without it being realized manifestly in the spectrum.
For example, such $SL(2,\mathbb{R})$ towers are not observed in the
numerically determined quasi-normal mode spectrum \cite{Leaver:1985ax}.

The conformal dimensions we find at lowest order in $M\omega$ are
given by \eqref{eq:condim}. It should be noted that the conformal
dimensions encountered are all positive, indicative of an underlying
unitary conformal field theory. 

When the higher order $M\omega$ terms in the Teukolsky equation are
included, the $SL(2,\mathbb{R})$ symmetry associated with the $L_{n}$'s
is softly broken. We expect the bulk scalar fields to be dual to CFT
operators involving a sum of higher dimension operators. The conformal
dimensions of these operators may be read off by examining the large
$r$ falloff of the expansion for the exact radial mode function \eqref{eq:expan},
yielding a prediction for the dimensions of other CFT operators that
must be present\begin{eqnarray*}
\Delta & = & -2|s|+2n+2\nu+2\,\qquad n>-\nu-1\\
 & = & 2|s|-2n-2\nu\,\qquad n<-\nu\end{eqnarray*}
which are again all positive.

In this way, each term in \eqref{eq:expan} can be interpreted as
a higher dimension correction in the mapping between the bulk mode
and CFT operators. Because \eqref{eq:expan} reproduces the exact
mode function for any finite $r$, one may deduce the exact two point
function for scattering of massless modes off Kerr, including all
higher $M\omega$ corrections, generalizing the lead-order matching
noted in \cite{Castro:2010fd}.

Having found a set of scaling dimensions associated with the exact
solution of the Teukolsky equation, we are confronted with the problem
that $\nu=\ell+\mathcal{O}(M^{2}\omega^{2})$ as shown in \eqref{eq:nueq}.
This means the scaling dimensions run with frequency -- another sign
that conformal symmetry is broken away from the $M\omega=0$ fixed
point. At first sight this seems rather disappointing: if we wish
to study the conformal fixed point, we are forced to set $M=0$. To
retain a smooth geometry, this limit must be taken with $a<M$ which
takes us to flat spacetime.%
\footnote{One may also consider the $M\to0$ limit with either fixed $a$ or
with fixed $J$. In each case, the limit of the $SL(2,\mathbb{R})\times SL(2,\mathbb{R})$
generators are not Killing vectors of the limiting metric, but rather
conformal symmetries of the massless field equations.%
} To keep the generators \eqref{eq:hgens} and \eqref{eq:hbars} well-defined,
one must also rescale the time coordinate, keeping $\tilde{t}=t/M$
finite and the dimensionless temperatures $T_{L}$ and $T_{R}$ fixed.
Thus the metric becomes $\mathbb{R}^{3}$ times a null direction $\tilde{t}$\begin{equation}
ds^{2}=0d\tilde{t}^{2}-dr^{2}-r^{2}d\theta^{2}-r^{2}\sin^{2}\theta\, d\phi^{2}\,.\label{eq:flatlimit}\end{equation}

Certainly we see no sign of a nontrivial central charge $c=12J$ associated
with an exact CFT dual to flat spacetime. One might have expected
extremal Kerr to emerge at the fixed point, but this simply does not
emerge in a small $M\omega$ limit. The generators \eqref{eq:hgens}
and \eqref{eq:hbars} are not isometries of the metric \eqref{eq:flatlimit}.
Moreover, in the extremal limit, they do not match the asymptotic
symmetry generators of the NHEK geometry found in \cite{Guica:2008mu,Matsuo:2009sj,Matsuo:2009pg}
(for further work in this direction see \cite{Matsuo:2010in,Chen:2010fr}).
Rather they correspond to conformal transformations of \eqref{eq:flatlimit}
that leave the massless wave equations invariant. Thus the hidden
Kerr/CFT correspondence does not seem easily generalized to massive
modes. 

Of course our scaling dimension computations are only valid at {}``strong''
coupling where the gravitational solution is smooth. There could still
be a nontrivial CFT with $c=12J$ with conformal dimensions that match
those obtained here when its strong coupling limit is taken. Studies
of the near super-radiant modes of extremal Kerr provide strong evidence
for such a conformal field theory \cite{Guica:2008mu,Bredberg:2009pv}.
While the two limits do not seem to be smoothly connected within the
realm of smooth gravity solutions (for example the super-radiant modes
do not satisfy the low frequency limit needed to obtain the symmetry
studied in the present paper), they may well be connected within the
exact microscopic CFT. Similar phenomena are observed in the duality
between D1,D5-brane backgrounds and CFT. 
\begin{acknowledgments}
I.M. thanks Cristian Vergu for helpful discussions. This research
is supported in part by DOE grant DE-FG02-91ER40688-Task A. 
\end{acknowledgments}
\bibliographystyle{brownphys}
\bibliography{hidden}

\end{document}